\documentclass{article}

\usepackage[dvips]{graphics,graphicx,color}
\usepackage{a4,indentfirst,latexsym,amsfonts,amsmath,amssymb,array}

\begin{document}

\title{Theoretical calculations of the primary defects induced by pions and protons in %%@
SiC}

\author {S. Lazanu\footnote{National Institute for Materials Physics, POBox MG-7, %%@
Bucharest-Magurele, Romania}, \hspace{1 pt}
I. Lazanu\footnote{University of Bucharest, POBox MG-11, Bucharest-Magurele, Romania}, %%@
\hspace{1 pt} 
E. Borchi\footnote{Universita di Firenze, Dipartimento di Energetica,
Via S. Marta 3, 50139 Firenze, Italy} \hspace{1 pt}
and M. Bruzzi\footnote{Universita di Firenze, Dipartimento di Energetica,
Via S. Marta 3, 50139 Firenze, Italy}} 
\date{\today}
\maketitle
\begin{abstract}

In the present work, the bulk degradation of SiC in hadron (pion and proton) fields, in %%@
the energy range between 100 MeV and 10 GeV, is characterised theoretically by means of %%@
the concentration of primary defects per unit fluence. The results are compared to the %%@
similar ones corresponding to diamond, silicon and GaAs.


61.82.-d: Radiation effects on specific materials.\\
\medskip
\end{abstract}

\section{Introduction. General properties}

Silicon carbide is a new interesting material for use in optoelectronic and electronic %%@
devices, detectors, for the use in nuclear waste technology or in fusion reactors. In all %%@
these applications, this semiconductor material could be exposed to extreme conditions of %%@
temperature, to chemical agents and to radiation fields.

SiC is not a simple semiconductor, it is a whole class of semiconductors
because of its polytypism \cite{1}. Polytypism refers to one-dimensional
polymorphism, i.e. the existence of different stackings of the basic
structural elements along one direction. More than 200 polytypes have been
reported \cite{2}. Only few of them have practical importance. These include
the cubic form $3C\left( \beta \right) $, and the $4H$ and $6H$ hexagonal
forms. For the cubic polytype, the symmetry group is $T_{d}^{2}$, while for
the hexagonal ones this is $C_{6v}^{4}$.

Some material characteristics of different polytypes of SiC, useful for its utilisation in %%@
radiation fields, are presented in comparison with the corresponding ones of diamond, %%@
silicon and GaAs in Table 1. The compilation of data has been  realised using Refs %%@
\cite{3,4,5,6,7,8,9,10,11,12}.

The listed material properties could explain why electronic devices made from SiC can %%@
operate at extremely high temperatures without suffering from intrinsic conduction effects %%@
due to its wide bandgap. Also, this characteristic allows to emit short wavelength light %%@
when fabricated into a light emitting diode structure. 
%\documentclass{article}

%\usepackage[dvips]{graphics,graphicx,color}
%\usepackage{a4,indentfirst,latexsym,amsfonts,amsmath,amssymb,array}

%\begin{document}

\begin{table}

\caption{}
\small 
\centering
\rotatebox{90}
{\begin{tabular}{|l|l|l|l|l|l|l|}
\hline
\textbf{Material properties} & \textbf{Diamond} & \textbf{Silicon} & \textbf{$6H$SiC} & %%@
\textbf{$4H$SiC}& \textbf{$3C$SiC}& \textbf{GaAs} \\ \hline
Crystal structure & Diamond \cite{3} & Diamond \cite{3} & Wurtzite \cite{3} & Wurtzite %%@
\cite{3} & Zincblende \cite{3} & Zincblende \cite{3} \\ 
Crystal system & Cubic \cite{3} & Cubic \cite{3} & Hexagonal \cite{3} & Hexagonal \cite{3} %%@
& Cubic \cite{3} & Cubic \cite{3} \\ 
Latice constant [A] & 3.567 \cite{3} & 5.431 \cite{3} & a: 3.08 \cite{3} & a: 3.09  %%@
\cite{3} & 4.38 \cite{3} & 5.85 \cite{3} \\ 
 &  &  & c: 15.12 \cite{3} & c: 10.8 \cite{3} \\ 
Atomic number & 6 & 14 & 6/14 & 6/14 & 6/14 & 31/33 \\ 
Atomic mass & 12.011 & 28.0855 & 12.01/28.08 & 12.01/28.08 & 12.01/28.08 & 69.7/74.9 \\
Density [g/cm$^3$] & 3.566 \cite{3} & 2.326 \cite{3}& 3.21 \cite{3} &  & 3.17 \cite{3} & %%@
5.32 \cite{3} \\ 
Radiation length [g/cm$^2$] & 42.7 \cite{4} & 21.82 \cite{4} & 28.9 &28.9 &28.9 & 12.2 \\ 
Mobilities & e: 1800 \cite{5} & e:1450 \cite{5} & e:450 \cite{3} & e:900 \cite{3} &e: 1000 %%@
\cite{3} & e: 8500 \cite{5} \\ 
 & h:1200 \cite{5} & h:450 \cite{5} & h:50 \cite{3} & h:100 \cite{3} & h:100 \cite{3} & %%@
h:400 \cite{5}\\ 
Band gap [eV]  & 5.47 \cite{5} & 1.12 \cite{5} & 3.02 \cite{3} & 3.26 \cite{3} & 2.2 %%@
\cite{3} & 1.43 \cite{5} \\ 
Displacement energy & 43 \cite{9}& 21 \cite{10} & Si:35 & Si:35 \cite{8} & Si:35 &10 %%@
\cite{10}\\ 
averaged value [eV] & & & C:20 & C:20 \cite{8} & C:20 \\
Dielectric constant & 5.5 \cite{6} & 11.8 \cite{6} & 9.61 \cite{6} & & 9.7 \cite{6} & 12.5 %%@
\cite{6}\\ 
Breakdown electric field &  & 2.5x10$^5$ & 2.4x10$^5$ & 2.2x10$^5$ & 2x10$^6$ & 3x10$^5$   %%@
\\ 
V/cm (for 1000 V operation) &  & & & & &  \\
Thermal conductivity & 20 \cite{6} & 1.5 \cite{6} & 4.9 \cite{6} &4.9 \cite{6} &5.0 %%@
\cite{6} & 0.48 \cite{6} \\
 @RT & & & & & & \\ 
Saturated electron drift velocity & 2.7x10$^7$ \cite{6} & 10$^7$ \cite{6} & 2x10$^7$ %%@
\cite{6} & 2.5x10$^7$ \cite{6} & 10$^7$ \cite{6}  \\ 
cms$^{-1}$@2x10$^5$ Vcm$^{-1}$&  &  & & & &  \\ 
dE/dx$_{min}$[MeVcm$^2$/g] & 1.78 \cite{5} & 1.664 \cite{5} & 1.35 &1.35 & 1.35 &1.37 %%@
\cite{5} \\ 
Energy to form an electron- & 13 \cite{7} & 3.72   & &8.4 \cite{7} & & 4.6 \cite{7}  \\ 
hole pair [eV] @ RT &  &  & & & &  \\ \hline
\end{tabular}
}
\end{table}
%\end{document}

The high breakdown electric field, approximately one order of magnitude higher than the %%@
silicon or GaAs ones, and the high thermal conductivity, enable the fabrication of very %%@
high-voltage, high-power devices that dissipate large amounts of the excess heat %%@
generated. Additionally, it allows the devices to be placed very close together, providing %%@
high device packing density for integrated circuits. SiC devices can also operate at high %%@
frequencies (microwave) because of the high saturation electron drift velocity in the %%@
material.

Some SiC material characteristics are useful for detection applications: its radiation %%@
length lies between the corresponding values of diamond and silicon; its relatively wide %%@
band gap ($2.2 \div 3.3$) eV and a relative dielectric constant of 7.04 allows the %%@
assembly of detectors with low noise. With low mobilities: 450 cm$^2$V$^{-1}$s$^{-1}$ for %%@
electrons and 115 cm$^2$V$^{-1}$s$^{-1}$ for holes respectively, SiC does not permit a %%@
fast signal development and therefore is not able to operate in high fluxes of particles.

The crystalline materials for semiconductor devices used in high fluences of particles are %%@
strongly affected by the effects of radiation. Up to now, in spite of the experimental and %%@
theoretical efforts, most of the problems related to the behaviour of semiconductor %%@
materials in radiation fields, the identification of the induced defects and their %%@
characterisation, as well as the explanation of the degradation mechanisms are still open %%@
problems.

Diamond has the reputation of being a radiation hard material and is considered as a  %%@
competitor to silicon. It is supposed that the concentration of primary defects induced in %%@
SiC in radiation fields is between the corresponding ones in diamond and silicon %%@
respectively.

\section{Mechanisms of degradation}

At the passage of the incident charged particle in the semiconductor material same of its %%@
energy is deposited into the target. The charged particles interact with both nuclei and %%@
electrons in a solid. The total rate of energy loss, could, in general, be divided %%@
artificially into two components, the nuclear and the electronic part.

The energy lost due to interactions with the electrons of the target gives rise to %%@
material ionisation, while the energy lost in interactions with nuclei is the origin of %%@
defect creation. The primary radiation defects are vacancies and interstitials.

There is not one physical quantity dedicated to the global characterisation of the effects %%@
of radiation in the semiconductor material. A possible one is the concentration of primary %%@
radiation induced defects per unit particle fluence ($CPD$) \cite{13}. It is not %%@
proportional to the modifications of material parameters due to the irradiation because of %%@
the time evolution of vacancies and interstitials - see the discussion in reference %%@
\cite{14}. Due to the fact that it does not exist a theory describing these processes of %%@
interaction, phenomenological models are used.

In this paper, the mechanism considered in the study of the interaction between the %%@
incoming particle and the solid, by which bulk defects are produced, is the following: the %%@
particle, heavier than the electron, with electrical charge or not, interacts with the %%@
electrons and with the nuclei of the crystalline lattice. The nuclear interaction produces %%@
bulk defects. As a result of the interaction, depending on the energy and on the nature of %%@
the incident particle, one or more light particles are produced, and usually one or more %%@
heavy recoil nuclei. These nuclei have charge and mass numbers lower or at least equal %%@
with those of the medium \cite{15}. After the interaction process, the recoil nucleus or %%@
nuclei, if they have sufficient energy, are displaced from the lattice positions into %%@
interstitials. Then, the primary knock-on nucleus, if its energy is large enough, can %%@
produce the displacement of a new nucleus, and the process could continue as a cascade, %%@
until the energy of the nucleus becomes lower than the threshold for atomic displacements. %%@
Because of the regular nature of the crystalline lattice, the displacement energy is %%@
anisotropic. The primary interaction between the hadron and the nucleus of the lattice %%@
presents characteristics reflecting the peculiarities of the hadron, especially at %%@
relatively low energies. If the inelastic process is initiated by nucleons, the identity %%@
of the incoming projectile is lost, and the creation of secondary particles is associated %%@
with energy exchanges which are of the order of MeV or larger. For pion nucleus processes, %%@
the absorption, the process by which the pion disappears as a real particle, is also %%@
possible.

The energy dependence of cross sections, for proton and pion interaction with the nucleus, %%@
present very different behaviours: the proton-nucleus cross sections \cite{16,17} decrease %%@
with the increase of the projectile energy, have a minimum at relatively low energies, in %%@
the range of 100 $\div$ 200 MeV, followed by a smooth increase, while the pion nucleus %%@
cross sections present for all processes a large maximum, at about 160 MeV, reflecting the %%@
resonant structure of interaction (the $\Delta_{33}$ resonance production), followed by %%@
other resonances, at higher energies, but with much less importance \cite{18}. Because of %%@
the multitude of inelastic channels, in the concrete evaluation of defect production, the %%@
nuclear interactions has been modelled previously in some simplifying hypothesis, see %%@
refs. \cite{13,18,19,20}.

The process of partitioning the energy of the recoil nuclei (produced due the interaction %%@
of the incident particle with the nucleus, placed in its lattice site) by new interaction %%@
processes, between electrons (ionisation) and atomic motion (displacements) is considered %%@
in the frame of the Lindhard theory \cite{21}.

\section{Results and discussions}

The concentration of the primary radiation induced defects per unit fluence in SiC has %%@
been calculated using the explicit formula:

\begin{equation}
\begin{split}
CPD	\left(E\right)&= \frac{N_{C}}{2E_{d;C}} \int \sum _{i} \left( \frac{d\sigma}{d\Omega} %%@
\right)_{i;C} L(E_{Ri})_{SiC} d\Omega
\\ 
&+\frac{N_{Si}}{2E_{d;Si}} \int \sum _{i} \left( \frac{d\sigma}{d\Omega} \right)_{i;Si} %%@
L(E_{Ri})_{SiC} d\Omega
\\
&=\frac{1}{N_A} \left[ %%@
\frac{N_CA_C}{2E_{d;C}}(NIEL)_{SiC}^{Cfamily}+\frac{N_CA_C}{2E_{d;Si}}(NIEL)_{SiC}^{Sifami%%@
%%@
ly} \right]
\end{split}
\end{equation}
where $E$ is the kinetic energy of the incident particle, $N_{C (Si)}$  is the atomic %%@
density of the $C (Si)$ in SiC, $A_{C (Si)}$ is the atomic number of the C (Si), $E_{d;C %%@
(Si)}$ - the threshold energy for displacements in the C (Si) sublattice of SiC, $E_{Ri}$ %%@
- the recoil energy of the residual nucleus produced in interaction $i$,  $L\left( %%@
E_{Ri}\right) $ - the Lindhard factor that describes the partition of the recoil energy %%@
between ionisation and displacements and $d\sigma _i/d\Omega $ - the differential cross %%@
section of the interaction between the incident particle and the nucleus of the lattice %%@
for the process or mechanism $i$, responsible in defect production. The atomic density %%@
$N_{C (Si)}$  in SiC is a material parameter which depends to the polytype through %%@
structural constants. The formula gives also the relation with the non ionising energy %%@
loss ($NIEL$). $N_A$ is the Avogadro number. It is important to observe that there exists %%@
a proportionality between the $CPD$ and $NIEL$ only for monoelement materials, and this %%@
proportionality can be extended only approximately for compounds with remote elements in %%@
the periodic table, e.g. for GaAs.

The $CPD$ allows the comparison of the effects produced by the same particle in different %%@
materials, while $NIEL$ is especially used for the comparison of the effects produced in %%@
the same material by different particles.

The partition of the energy of the incoming particle in the semiconductor material between %%@
ionisation and displacements has been calculated analytically, solving the general %%@
equations of the Lindhard theory in some physical approximations. Details about the %%@
hypothesis used could be found in reference \cite{22}. In the particular case of the SiC, %%@
a binary compound, the Lindhard curves have been calculated separately for each component, %%@
and the average weight Bragg additivity has been used. The energy spent into displacements %%@
is represented versus recoil energy for different recoils, characterised by their charge %%@
and mass numbers, in SiC, in Figure 1.

\begin{figure}[ht]
\centering
\includegraphics[width=.8\textwidth,clip,angle=90]{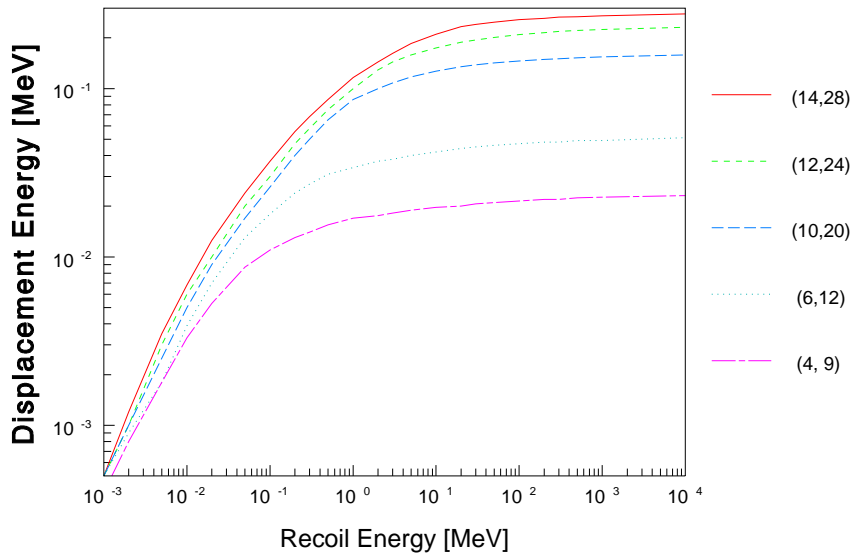}
\caption{\small{The displacement energy versus the energy of recoils (characterised by %%@
their atomic and mass number) in SiC.}}
\label{f1}
\end{figure}

All curves start, at low energies, from the same curve; they have at low energies %%@
identical values of the energy spent into displacements, independent on the charge and %%@
mass number of the recoil. At higher energies, the curves start to detach from this main %%@
branch. This happens at lower energies if their charge and mass numbers are smaller. The %%@
maximum energy transferred into displacements corresponds to recoils of maximum possible %%@
charge and mass numbers (corresponding to the heaviest element, Si in this case). The %%@
curves present then a smooth increase with the energy. For the energy range considered %%@
here, the asymptotic limit of the displacement energy is not reached.

For the cross sections for proton and pion interactions with carbon and silicon, %%@
experimental data and theoretical extrapolations have been used - see references %%@
\cite{19,23,24} and references cited therein. The cross sections for pion Ga and As nuclei %%@
are not available from experiment. Due to this fact, parametrisations of the energy %%@
dependence of the cross sections  have been tried. Details about the procedure are given %%@
in reference \cite{25}.

As stated in Section 2, the minimum energy required to form a stable vacancy - %%@
interstitial pair is usually called the threshold energy for displacements ($E_d$). The %%@
regular nature of the crystalline lattice conduces to its anisotropy, atoms being more %%@
readily displaced along some directions than in others. This anisotropy is also the origin %%@
of differences in the rate production of defects as a function of orientation. In the %%@
literature, both calculated and experimental values for   are reported for different %%@
crystals.

In principle, for SiC, the situation is even more complicated and a threshold for atomic %%@
displacements could be defined for each polytype, for each atomic species, and for each %%@
direction in the crystal. 

Some calculations have been reported in the literature for $E_d$: in SiC, along some %%@
directions of interest, by classical molecular-dynamics simulations using empirical %%@
potentials \cite{26}, and also using modern first-principle methods \cite{27}.
 
Experimental values are obtained from high voltage electron microscopy \cite{28}, %%@
energetic ion beam spectroscopy \cite{29}, and Rutherford backscattering analysis along %%@
different channelling directions \cite{10,29,30}.
 
In the opinion of some authors \cite{30}, all these methods give consistent results, the %%@
recommended mean threshold energy for atomic displacements being of the order of 35 eV on %%@
the Si sublattice, and of the order of 20 eV on the C sublattice. These are the values we %%@
used in the calculations.

The energy dependence of the $CPD$ calculated (as stated above) for pions in SiC, together %%@
with the estimated values for proton irradiation is presented in Figure 2. In the same %%@
figure, for comparison, the $CPD$ produced by pions and protons in diamond \cite{19,24}, %%@
silicon and GaAs are also represented. The values corresponding to protons in silicon are %%@
averages from the calculations reported in references \cite{12} and \cite{31}. The curve %%@
for protons in GaAs is obtained from the results reported in reference \cite{12} and is %%@
estimated by extrapolation in the energy range of interest for the present study, while %%@
the curve for pions in GaAs is from reference \cite{23}. It could be observed that the %%@
degradations induced by pions and protons are very different and reflect the peculiarities %%@
of the interactions of the two particles with the nuclei of the semiconductor. For pions, %%@
the maximum of degradation is produced in the region of the $\Delta_{33}$ resonance, %%@
corresponding to about 140¸ 160 MeV kinetic energy. In comparison with this behaviour, the %%@
maximum of the degradation is produced, for protons, at low energies, and a smooth and %%@
slow decrease of the $CPD$ is calculated with the increase of the energy. 
\begin{figure}[ht]
\centering
\includegraphics[width=.8\textwidth,clip,angle=90]{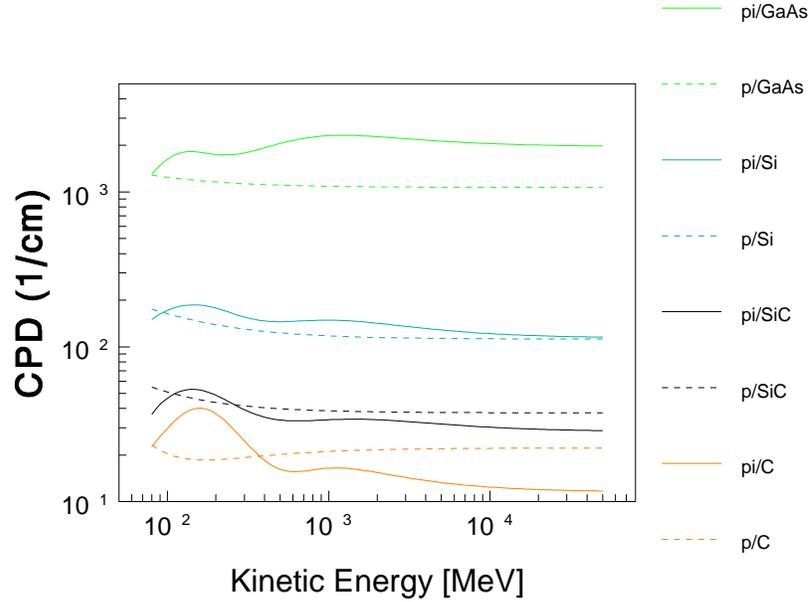}
\caption{\small{Concentration of primary induced defects on particle unit fluence for %%@
pions (continuous) and protons 
dashed) in (up to down): gallium arsenide,  silicon, silicon carbide and diamond.}}
\label{f2}
\end{figure}
For the energy range considered in this paper, with the exception of the resonance region, %%@
the degradation induced by protons is higher than that for pions in SiC, but for both %%@
incident particles, the concentration of primary defects produced per unit fluence are %%@
closed between 30 and 40 cm$^{-1}$. The degradation induced by hadrons in SiC is between %%@
the diamond and silicon degradations.

Also, it could be observed that the pion degradation becomes more important with the %%@
increase of the mass number of the material, in the whole range of energy investigated.

The $CPD$ for $3C$, $4H$ and $6H$ polytypes of SiC has been evaluated separately, %%@
considering the structural characteristics of each polytype, and the same displacement %%@
energies for all of them; no significant differences between their behaviour in radiation %%@
fields has been found. These theoretical results are in agreement with the experimental %%@
data reported in reference \cite{32}.

\section{Summary}

In the present paper the concentration of primary induced defects per unit particle %%@
fluence has been calculated for the case of pion and proton irradiation, and the effects %%@
of these particles have been found intermediate in SiC between the corresponding ones in %%@
Si and diamond.

Negligible differences have been found between the $3C$, $4H$ and $6H$ polytypes in what %%@
regards the effects of pion and proton irradiation.

The Lindhard curves characterising the partition of the recoils energy between ionisation %%@
and displacements have been calculated using analytical approximations.

The main properties of SiC useful for its utilisation in radiation fields have been %%@
reviewed.

\end{document}